\documentclass[12pt]{reportj}
\usepackage{deluxetablej}
\usepackage{hyperref} 
\makeatletter
\def\@to{to}
\makeatother
\usepackage{times}
\usepackage{graphicx}
\usepackage{tocloft}
\usepackage{fancyheadings}
\emergencystretch=\maxdimen
\usepackage{datetime}
\newdateformat{ddmonthyyyy}{\THEDAY\ \monthname[\THEMONTH]\ \THEYEAR}

\renewcommand\thesection{\arabic{section}}
\renewcommand\thesubsection{\thesection.\arabic{subsection}}

\def\ssection#1{\setcounter{subsection}{0} \refstepcounter{section} \section*{\hbox to \hsize{\large\bf \arabic{section}. #1\hfill }}\label{sec} \addcontentsline{toc}{section}{\arabic{section}. #1}}
\def\ssubsection#1{\setcounter{subsubsection}{0} \refstepcounter{subsection}\subsection*{\hbox to \hsize{\normalsize\bfseries\itshape \arabic{section}.\arabic{subsection} #1\hfill}}\label{subsec} \addcontentsline{toc}{subsection}{\arabic{section}.\arabic{subsection} #1}}
\def\ssubsubsection#1{\refstepcounter{subsubsection}\subsection*{\hbox to \hsize{\normalsize\it \arabic{section}.\arabic{subsection}.\arabic{subsubsection} #1\hfill}}\label{subsubsec} \addcontentsline{toc}{subsubsection}{\arabic{section}.\arabic{subsection}.\arabic{subsubsection} #1}}

\def\ssectionstar#1{\section*{\hbox to \hsize{\large\bf #1\hfill}} \addcontentsline{toc}{section}{#1}}
\def\ssubsectionstar#1{\subsection*{\hbox to \hsize{\normalsize\bfseries\itshape #1\hfill}} \addcontentsline{toc}{subsection}{#1}}
\def\ssubsubsectionstar#1{\subsection*{\hbox to \hsize{\normalsize\it  #1\hfill}} \addcontentsline{toc}{subsection}{#1}}

\renewcommand{\cftaftertoctitle}{%
\mbox{}\hfill{\normalfont Page}}
\begin{document}

~\\

\vspace{-2.4cm}
\noindent\includegraphics*[width=0.295\linewidth]{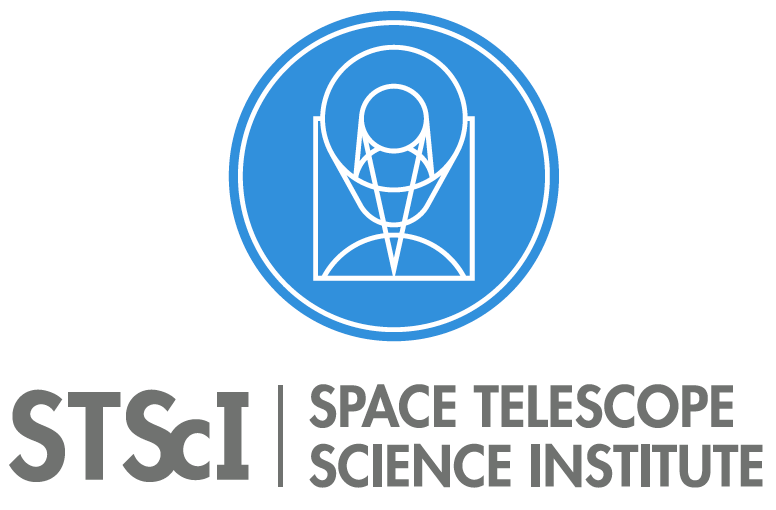}

\vspace{-0.4cm}

\begin{flushright}
 {\bf Instrument Science Report COS 2026-04(v1)}
 
 \vspace{1.1cm}
 
 {\bf\Huge COS2035: Extending COS/FUV Operations Through the 2030s}
 
 \rule{0.25\linewidth}{0.5pt}
 
 \vspace{0.5cm}
 
 Marc Rafelski$^1$, David Sahnow$^1$, Christian I. Johnson$^1$,  Bethan James$^1$, Svea Hernandez$^{1,2}$, John Debes$^{1,2}$, Beverly Serrano$^3$, Kate Davis$^1$, Serge Dieterich$^1$, Van Dixon$^1$, Leonardo Dos Santos$^1$, Travis Fischer$^{1,2}$, Elaine Frazer$^1$, David French$^1$, Mark Giuliano$^1$, Joshua Goldberg$^1$, Sten Hasselquist$^1$, Jacqueline Hernandez$^1$, Nick Indriolo$^{1,2}$, Mike Kelly$^3$, Olivia Lupie$^3$, Lauren Miller$^1$, Diego Mundo$^1$, Anna Payne$^1$, Karla Peterson$^1$, Rachel Plesha$^1$, Kate Rowlands$^{1,2}$, Julia Roman-Duval$^1$, Ravi Sankrit$^1$, Debopam Som$^1$, Scott Swain$^3$, Alan Welty$^1$
 \linebreak
 \newline
 \footnotesize{$^1$ Space Telescope Science Institute, Baltimore, MD\\}
 \footnotesize{$^2$ AURA for the European Space Agency\\}
 \footnotesize{$^3$ NASA Goddard}

 \vspace{0.2cm}
 
13 July 2026
\end{flushright}

\vspace{0.1cm}

\noindent\rule{\linewidth}{1.0pt}
\noindent{\bf A{\footnotesize BSTRACT}}

{\it \noindent The far-ultraviolet (FUV) detector of the Cosmic Origins Spectrograph (COS) accumulates gain sag where photons land, and without continued mitigation this degradation would render the most used modes unusable. To extend COS FUV operations through the 2030s, the COS team developed the COS2035 strategy, which builds on the existing COS2025 rules with four technical breakthroughs and two new usage policies. First, SPLIT-wavecals decouple wavelength calibration from science exposures and open detector real estate above the Pt-Ne lamp light leak. Second, a hybrid lifetime position (LP) architecture allows different gratings to operate at different LPs simultaneously. Third, the LP-infinity framework removes the dependence on the eight-LP limit in the COS flight software, supported by a new table-based APT and TRANS rules architecture. Fourth, a revised gain-sag flagging method evaluates integrated column count loss against the maximum achievable signal-to-noise (S/N) per mode. The two new usage policies cap per-target S/N at the maximum achievable value set by fixed-pattern noise, and limit any single program to 2\% of the lifetime at any single LP. With LP7 and LP10 enabled in Cycle 33 and LP11 and LP12 in active commissioning for Cycles 34 and 35, the COS2035 strategy positions the FUV channel for continued high productivity through the 2030s.}

\vspace{-0.1cm}
\noindent\rule{\linewidth}{1.0pt}

\newpage

\lhead{}
\rhead{}
\cfoot{\rm {\hspace{-1.9cm} Instrument Science Report COS 2026-04(v1) Page \thepage}}

\renewcommand{\cftaftertoctitle}{\thispagestyle{fancy}}
\tableofcontents


\vspace{-0.3cm}
\ssection{Introduction}\label{sec:intro}

The Cosmic Origins Spectrograph (COS) was installed on the Hubble Space Telescope (HST) during Servicing Mission 4 in May 2009 and has since obtained tens of thousands of ultraviolet spectra, predominantly with the far-ultraviolet (FUV) channel. The FUV channel uses a cross-delay-line microchannel plate detector operated at high voltage, and prolonged exposure to light causes the gain to sag in regions of the detector where photons accumulate. As the gain falls, the centroiding of photon events becomes unreliable and the apparent sensitivity drops in the affected columns. To preserve data quality, the COS team periodically commissions new lifetime positions (LPs) by relocating the science spectrum to regions of the detector with adequate gain. Between 2009 and 2018, four lifetime positions (LP1 through LP4) were commissioned in the central detector region accessible at the time, with all FUV gratings and both science apertures sharing a single LP at any given time (Osten et al. 2013, LP2; Roman-Duval et al. 2016, LP3; Roman-Duval et al. 2018, LP4). The instrument was designed to support up to five lifetime positions, with the FUV aperture plate sized to provide internal flat-field coverage across that range (Morse 1999).

\begin{figure}[!b]
  \centering
  \includegraphics[width=\textwidth]{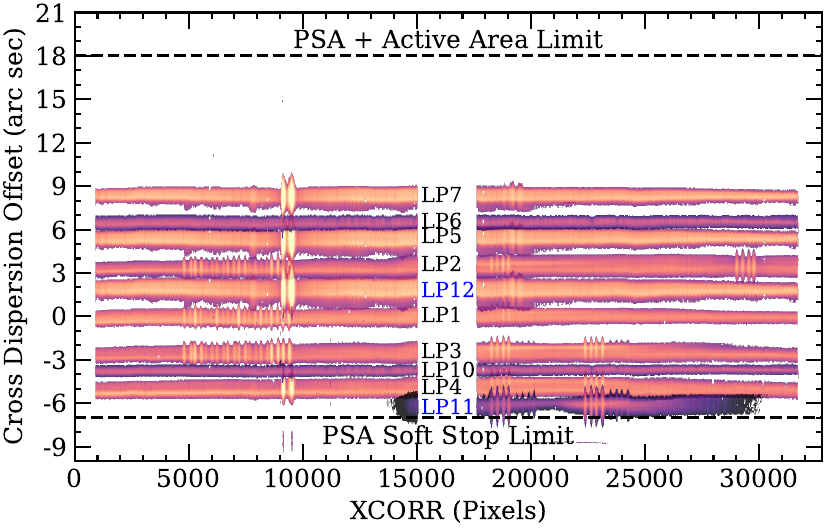}
 \caption{Cross-dispersion locations of the COS FUV lifetime positions across the two detector segments, shown as the accumulated two-dimensional spectral traces as a function of XCORR pixel and cross-dispersion offset from LP1. The dashed lines mark the PSA active-area limit at the top of the usable region and the PSA soft stop at the bottom. LP1 through LP7 and LP10 are commissioned positions; LP11 and LP12 (blue) are in commissioning for Cycles 34 and 35. The operational boundaries that constrain lifetime position placement, including the light leak, FCA illumination range, and aperture mechanism soft stops, are shown in Figure \ref{fig:realestate}.}
 \label{fig:lpmap}
\end{figure}

Despite the presence of considerable detector area above and below the central lifetime positions, only the central region of the detector had been considered viable for science. The cross-dispersion locations of all COS lifetime positions on the two detector segments are shown in Figure \ref{fig:lpmap}. Operations above LP5 (offset from LP1 by $+5.4''$) were limited by a light leak through the Flat-field Calibration Aperture (FCA) at aperture block positions corresponding to target positions greater than $+5.5''$ relative to LP1, which prevented the use of concurrent science and Pt-Ne lamp wavecal exposures (Oliveira et al. 2013). The same region of the detector also lay above the reach of the FCA illumination, so internal gain maps using the deuterium lamp through the FCA were not available there, and it sat beyond the allowed aperture mechanism range for the Bright Object Aperture (BOA). The detector area below the previous lifetime positions was constrained by the aperture mechanism soft stop, and the lower portion of the accessible region had not been calibrated for geometric distortion and walk corrections.

\begin{figure}[!h]
  \centering
  \includegraphics[width=0.7\textwidth, trim=2cm 2.2cm 14cm 2.4cm, clip]{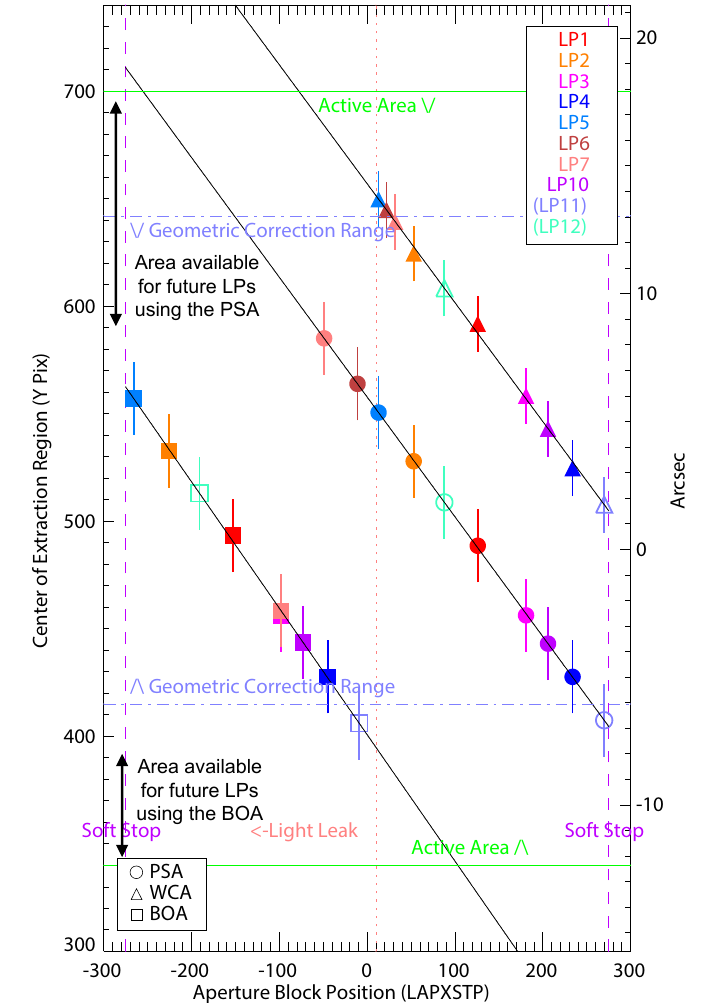}
 \caption{The COS FUV detector real estate showing the locations of the PSA, BOA, and WCA at each lifetime position, in relation to the operational boundaries (light leak, aperture mechanism soft stops, and the range over which geometric distortion corrections are available). For lifetime positions that use concurrent TAGFLASH wavecals, the WCA shares the same dispersion (X) position as the PSA and is offset only in cross-dispersion (Y), so the two spectra fall on different rows simultaneously. The BOA for a given LP maps to the same cross-dispersion (Y) location as that LP's PSA. The decoupling of the PSA, BOA, and WCA under the COS2035 strategy opens detector area up to the PSA active-area limit near $+18''$ from LP1 for PSA-only lifetime positions. LP11 and LP12, shown in parentheses and as open shapes, are in commissioning for Cycles 34 and 35.}
 \label{fig:realestate}
\end{figure}

Four related ideas, developed by the COS team in 2020, opened access to the rest of the detector for PSA science and laid the foundation of the COS2035 strategy. First, the COS team established that the aperture mechanism is sufficiently stable to be repositioned between exposures, enabling SPLIT-wavecals in which wavelength calibration exposures are taken at a separate aperture block position rather than concurrently with the science exposure (James et al. 2023a). Combined with a simulated 600 s lampflash that limits per-orbit overhead of SPLIT-wavecals (Section \ref{subsec:splitwavemethod}), this removes the light leak as a hard constraint on PSA placement (Rowlands et al. 2024). Second, the team recognized that the detector area above the $+5.5''$ light leak, previously used only by the WCA and considered unsuitable for science, is in fact nearly pristine and can host the PSA when SPLIT-wavecals are used (James et al. 2023b; Figure \ref{fig:realestate}). High aperture placement positions come at a cost, however, since spectral resolution decreases with distance from LP1 and operation above the light leak incurs the SPLIT-wavecal overhead. Third, the team recognized that different FUV modes can be placed at different LPs simultaneously in a hybrid-LP architecture, with each mode at the location best suited to its cross-dispersion profile, usage, and overhead sensitivity (James et al. 2023b). Fourth, the team recognized that the BOA need not be located at the same detector position as the PSA, as had previously been assumed, but can be decoupled from it, removing the requirement that the BOA be supported at every LP, which is rarely used, and freeing the PSA to move higher on the detector (Sahnow et al. 2021).

Together these four ideas allowed G130M to be placed at LP5 below the light leak and above LP2, and G160M long exposures to be placed at LP6 above LP5, each in a location chosen for its own properties (James et al. 2023a, James et al. 2023b). LP6 was placed as close to LP5 as the gain modeling allowed, both to preserve detector area higher up for additional LPs and to keep LP6 and those future positions as close to LP1 as possible, since spectral resolution decreases with distance from LP1. The available lifetime at LP6 proved to be more sensitive to LP5 usage than originally anticipated, and subsequent LPs have been placed with greater separation from their neighbors so that their lifetimes evolve more independently.

The recognition that different modes can be placed at different LPs has continued to drive the COS2035 strategy. As of the start of Cycle 33 in November 2025, the G130M cenwaves are split across two LPs. The bluer cenwaves (1055, 1096, 1222, and 1291) now sit at LP7 above the light leak and use SPLIT-wavecals; 1055 and 1096 had previously been at LP2. The redder cenwaves (1300, 1309, 1318, and 1327) remain available at LP5, but only for Segment A observations under the COS2025 rules (Oliveira et al. 2018). G160M, which has a substantially narrower cross-dispersion profile than G130M, was consolidated entirely at LP10 between LP3 and LP4, in a region that had previously been considered too gain-sagged for use. Placing G160M at LP10 recovers slightly higher spectral resolution than would have been available at higher offsets from LP1 and avoids the overhead penalty associated with SPLIT-wavecals, since LP10 sits below the light leak and uses concurrent wavecals.

G140L will be placed below LP4 at LP11 in Cycle 34 starting November 2026, leaving higher-resolution detector area available for other gratings. The G140L resolution is low across all LPs, and the resolution at LP11 is acceptable for that mode. LP11 will sit below the light leak and uses concurrent wavecals (TAGFLASH). In Cycle 35, G130M/1291, the most heavily used FUV mode, will be moved between LP1 and LP2 at LP12. This region carries residual gain sag from earlier LP1 and LP2 operations, but that sag is concentrated in the outer wings of the LP12 cross-dispersion profile. The revised gain-sag flagging method (Section \ref{sec:gainsag}), which evaluates integrated column count loss along a column against the maximum achievable signal-to-noise (S/N) of the mode rather than flagging individual pixels below modal gain 3, confirms that the residual sag does not significantly affect the integrated flux. LP12 will sit below the light leak and uses concurrent wavecals. This method, together with the LP-infinity ground system framework that removes the eight-slot LP limit in the COS flight software, rounds out the four technical breakthroughs of the COS2035 strategy.

Throughout this report, ``COS2035'' refers to the combined operational strategy, while the two usage policies are specifically the maximum-S/N cap and the per-program lifetime-usage limit. The remainder of this report describes the COS2035 strategy in detail. Section \ref{sec:splitwave} develops SPLIT-wavecals and the new detector real estate, including the geometric and operational consequences of decoupling the PSA, BOA, and WCA. Section \ref{sec:hybridlp} lays out the hybrid-LP architecture and the current grating and cenwave assignments across the detector. Section \ref{sec:framework} describes the LP-infinity framework and the associated APT and TRANS rules architecture that made adding new LPs a routine update rather than a major development effort. Section \ref{sec:gainsag} describes the revised gain-sag flagging method in detail. Section \ref{sec:strategy} synthesizes the strategy, layering the four technical breakthroughs with the COS2025 rules (Oliveira et al. 2018) and two new usage policies that limit per-target S/N and per-program lifetime usage. Section \ref{sec:forward} looks forward to LP11 and LP12 commissioning and to candidate locations for additional LPs. Section \ref{sec:summary} summarizes.

\vspace{-0.3cm}
\ssection{SPLIT-wavecals and the new detector real estate}\label{sec:splitwave}

Until 2020, three constraints were thought to bound the detector area above the previous lifetime positions, limiting COS operations to the central region of the detector. The light leak above approximately $+5.5''$ from LP1 prevents concurrent wavelength calibration with the WCA; the aperture mechanism soft stop prevents the BOA from reaching above approximately $+5.8''$; and the FCA illumination range, which extends to approximately the LP6 position, prevents internal gain maps from being obtained higher up. The COS2035 strategy works around each of these three constraints. SPLIT-wavecals decouple the WCA from the PSA in time, allowing wavelength calibration exposures to be taken at a different aperture block position than the science exposure, at the cost of higher overheads (approximately 15 to 30\% per orbit, with the lower end for two-exposure-per-orbit programs; Section \ref{subsec:overheads}). A revised BOA policy decouples the BOA from the PSA in detector position, removing the requirement that BOA observations be supported at every LP. External white dwarf observations replace the internal deuterium lamp for gain monitoring above the FCA illumination limit. Together these three approaches open the detector area above the previous lifetime positions for science, up to the PSA active-area limit near $+18''$ from LP1. Figure \ref{fig:realestate} shows how the PSA, BOA, and WCA are assigned to different regions of the detector under these changes.

\vspace{-0.3cm}
\ssubsection{SPLIT-wavecals and the simulated 600 s flash}\label{subsec:splitwavemethod}

Under TAGFLASH, the standard wavelength calibration mode used at LPs below the $+5.5''$ light leak, the WCA is illuminated concurrently with the PSA: the aperture block is fixed at a single position throughout the science exposure, the WCA transmits Pt-Ne lamp light at intervals during the integration, and the resulting wavecal lamp flashes are used to track and correct Optics Select Mechanism (OSM) drift in the wavelength solution. This works because the WCA sits at a fixed offset from the PSA in the cross-dispersion direction, so the two spectra fall on different rows of the detector and can be obtained simultaneously (Figure \ref{fig:realestate}). Above $+5.5''$, the high aperture block position required to place the PSA and WCA there also moves the FCA past the point where its mask blocks stray Pt-Ne light, so concurrent wavecals would shut the detector down due to count-rate violation.

SPLIT-wavecals replace concurrent illumination with sequential illumination: the aperture block is moved from the science position to a separate WCA-only position before each lamp flash, the lamp is flashed safely at that position, and the block is returned to the science position before the next science integration resumes. This sequence is repeated as the timing rules require, with one wavecal at the start of each science exposure and a second at the end of any exposure long enough to need it (James et al. 2023b). The aperture mechanism is sufficiently stable under repeated repositioning that the wavelength solutions obtained at the WCA-only position transfer to the science exposure with sub-pixel accuracy (Fox et al. 2020).

For exposures longer than 960 s, the standard TAGFLASH timing rules call for an additional middle wavecal at 600 s after a large OSM motion, such as a grating change at the start of an exposure, to track the OSM drift that is most pronounced early in an exposure. Carrying that middle flash through SPLIT-wavecals would have added significantly to the per-exposure overhead, and the COS team developed a method for removing it without significantly degrading the wavelength solution. The drift at 600 s is simulated from the drift measured at the end of the exposure using a piecewise function fit to archival COS data, with parameters depending on grating, exposure time, and the size of the end-of-exposure drift (Rowlands et al. 2024; Dos Santos et al. 2025). The simulated 600 s flash recovers the wavelength solution in the affected portion of the exposure to within roughly 0.5 pixels for approximately 85--90\% of exposures, well within the COS wavelength calibration error budget of $\pm 3$ pixels in the dispersion direction (Plesha et al. 2018). The simulation is implemented in CalCOS via reference parameters in the WCPTAB and is applied automatically to SPLIT-wavecal exposures meeting the timing criteria.

\vspace{-0.3cm}
\ssubsection{BOA decoupling and the new PSA-only real estate}\label{subsec:boa}

The PSA, WCA, and BOA occupy fixed positions on the aperture block. The WCA sits above the PSA in cross-dispersion, so during concurrent TAGFLASH observations the wavecal spectrum projects a fixed number of Y pixels above the science spectrum on the detector. The BOA, by contrast, sits below the PSA on the aperture block, but lifetime positions are defined by sky position rather than detector location. Observing a given LP through the BOA requires moving the aperture block by 279 steps (LAPXSTP) so that the BOA replaces the PSA at the HST pointing. The BOA spectrum therefore lands at approximately the same cross-dispersion location on the detector as that LP's PSA spectrum, which is how the two appear in Figure \ref{fig:realestate}, but it is reached at a different aperture-block position. Soft stops at $\pm 275$ LAPXSTP limit the total range of motion, precluding BOA use above approximately $+5.8''$ relative to LP1.

For the first decade of COS operations, BOA support was treated as a requirement at every LP. This was the constraint that, together with the WCA light leak above $+5.5''$, limited the usable detector area to the central region. BOA observations are very uncommon: only a single FUV program has ever used the BOA for COS science, and BOA support has been maintained at a small number of LPs without the full calibration program applied to PSA observations. Bright-target science cases that might use the BOA are generally better served by the STIS spectrograph. The COS team's decision to relax the BOA-at-every-LP requirement was one piece of the strategic shift that opened access to the rest of the detector. Under current operations, G130M and G160M BOA observations are placed at LP4 and G140L BOA observations are placed at LP3, with BOA support not extended to LP5 or higher. If STIS becomes unavailable in the future, the COS team can commission new BOA-supporting LPs for bright targets. We note that because the BOA sits below the PSA, a BOA-supporting LP placed low on the detector would need external-target gain monitoring, since the FCA becomes vignetted at that aperture block position.

Removing the requirement that the BOA be accessible at every LP opened access to a substantial new region of the detector for PSA-only lifetime positions, extending upward toward the PSA active-area limit near $+18''$ from LP1. This region had previously been illuminated only by the WCA and was known to be safe, but the resolution, focus, and geometric distortion at large offsets from LP1 had not been explored at the time of the LP5 Exploratory Study (James et al. 2023a). The LP6 Exploratory Study (James et al. 2023b) characterized these properties up to $+11''$ and mapped the modal gain out to $+13''$ where it was found to remain well above the operational threshold. The detector area above $+13''$ has not yet been characterized but is not known to carry any limitation on lifetime position placement. The study established LP6 at $+6.5''$ as the optimal placement for the first PSA-only LP, balancing resolution against the gain-sag interaction with the LP5 spectral profile while preserving detector area higher up for additional LPs. Under the COS2035 strategy, subsequent LPs in the new real estate, including LP7 at $+8.3''$ hosting most G130M cenwaves and future lifetime positions above LP7, follow the same logic, with each placement chosen to balance resolution against neighbor-LP interactions.

\vspace{-0.3cm}
\ssubsection{Overheads and supporting infrastructure}\label{subsec:overheads}

The overhead consequences of SPLIT-wavecals depend on how many science exposures fit in an orbit. Without the simulated 600 s flash, SPLIT-wavecal overheads added 24--30\% per orbit to typical observing patterns (James et al. 2023b). With the 600 s flash removed, the overhead increase falls to approximately 15\% for two exposures per orbit, with no reduction for four-exposure-per-orbit programs since those exposures are too short to trigger the middle flash even under TAGFLASH (James et al. 2023a). These overhead numbers shape how SPLIT-wavecal LPs are used in the broader COS2035 architecture, with different exposure profiles routed to different LPs as described in Section \ref{sec:hybridlp}.

The WCA real estate is itself a finite resource that must be partitioned across multiple SPLIT-wavecal LPs. These WCA aperture definitions are set in the ground system keyword rules rather than by any hardware limit, so they can be revised at any time and apply to data from previous observations when those data are recalibrated. The definitions used through LP4 occupied 20 steps each in aperture mechanism position. Tightening the LP2 WCA to 10 steps and adopting 10-step widths for the new SPLIT-wavecal WCAs created room for three additional WCA positions between LP2 and LP5 (James et al. 2023b); LP6 and LP7 each use one of these three slots, leaving one available for a future SPLIT-wavecal LP. Additional WCA real estate can be freed by shrinking the still-20-step WCA definitions of LP1, LP3, and LP4 to 10 steps as needed. The observed aperture distributions are only about 5 steps wide (James et al. 2023b), so the definitions could be narrowed further still if additional WCA positions are required.

Gain monitoring above the FCA illumination limit is performed using external white dwarf observations in place of the internal deuterium lamp. A dedicated calibration program observes WD-0308-565 with G130M/1222 and G160M/1623 across two orbits at each new LP, providing sufficient counts over the full segment area to construct gain maps comparable in quality to the internal FCA-illuminated maps used for the central LPs. Together with the BOA decoupling described in Section \ref{subsec:boa} and the SPLIT-wavecal capability described in Section \ref{subsec:splitwavemethod}, this external gain monitoring infrastructure makes the new detector real estate a generalizable resource for the COS2035 strategy rather than a one-time adaptation. This methodology currently supports operations at LP7 and will extend to all future LPs above LP7.

\vspace{-0.3cm}
\ssection{Hybrid-LP architecture}\label{sec:hybridlp}

The COS2035 strategy treats lifetime position assignment as an optimization across modes rather than a single-position choice for the instrument as a whole. Each FUV grating and cenwave is placed at the LP best suited to its cross-dispersion profile based on existing gain sag conditions, its expected usage, and its sensitivity to the SPLIT-wavecal overhead penalty. The result is a hybrid architecture in which different gratings or cenwave configurations operate at different LPs simultaneously, each at the high-voltage setting appropriate to the detector condition at its location, a change from earlier operations in which only one LP was active at a time. The plan will be revisited each cycle as detector conditions and usage patterns evolve. The current LP placements are shown in Figure \ref{fig:lpmap}.

Two principles govern the allocation of detector real estate. First, narrower cross-dispersion profiles can fit between previous LPs that wider modes cannot share. Second, assigning gratings to different LPs reduces the cross-dispersion footprint each LP requires. G160M, the narrowest of the FUV medium-resolution profiles, illustrates the first: it fits between LP3 and LP4 at LP10 (Figure \ref{fig:lpmap}), achieving higher spectral resolution than at higher offsets from LP1 and avoiding the SPLIT-wavecal overhead since LP10 sits below the $+5.5''$ light leak. The second arises because different gratings project to slightly different cross-dispersion positions even at the same nominal LP, due to slight misalignments between them; a single LP hosting multiple gratings must span the union of those projections, while a hybrid LP hosting one grating spans only its own profile. G130M/1291 fits between LP1 and LP2 at LP12 because of this narrower footprint, in combination with the revised gain-sag flagging method (Section \ref{sec:gainsag}).

Two further principles concern the interaction between LP placement and mode usage. First, the separation between nearby LPs affects lifetime independence. Second, mode placement should match the typical observing pattern. The LP5 and LP6 placements (James et al. 2023b) illustrate the first: they were placed as close as gain modeling permitted to preserve detector area higher up, but the resulting proximity meant high G130M usage at LP5 propagated into reduced lifetime at LP6. Subsequent LPs have been placed with greater separation, with LP7 at $+8.3''$ sitting well above LP6 at $+6.5''$ so the two evolve more independently. G140L illustrates the second: its observations are typically single-orbit pointings of bright targets with four exposures per orbit, a pattern that significantly benefits from the absence of SPLIT-wavecal overhead and thus motivates placement at LP11 below the light leak in Cycle 34. The blue G130M cenwaves (1055 and 1096), by contrast, have lower throughputs and therefore each exposure spans most of an orbit, so the SPLIT-wavecal overhead at LP7 is proportionally small.

The current grating and cenwave assignments under this architecture, including the LP11 and LP12 commissioning planned for Cycles 34 and 35, are summarized in Table \ref{tab:assignments}. The G130M, G160M, and G140L assignments are described in turn below.

The G130M cenwaves are split across two LPs as of the Cycle 33 transition. The bluer cenwaves (1055, 1096, 1222, and 1291) were moved to LP7 with SPLIT-wavecal in November 2025 after the gain at their previous positions (LP2 and LP5) had declined to the point that continued operation there was no longer viable. The redder cenwaves (1300, 1309, 1318, and 1327) remain at LP5 in Segment A only under the COS2025 rules, where Segment A still has adequate gain to support continued operations and TAGFLASH allows short exposures without an overhead penalty. G130M/1291 is planned to migrate from LP7 to LP12 in Cycle 35. Because G130M/1291 accounts for approximately 40\% of FUV usage by exposure count (based on Cycle 33 data), the placement of this single mode has substantial impact on the overall productivity of the FUV channel. LP12 will sit between LP1 and LP2 and provides the highest spectral resolution for science since LP1, without the SPLIT-wavecal overhead penalty associated with LP7.

G160M was consolidated entirely at LP10 in November 2025, with both long and short exposures now hosted at the same LP. The previous practice of placing G160M short exposures at LP4 alongside long exposures at LP6 ended with the LP10 transition. With LP10 carrying no SPLIT-wavecal overhead penalty, the historical motivation for the split-by-exposure-length placement no longer applies, and the gain at LP4 was running low. As of Cycle 33, the COS2035 architecture does not currently include any separation of long and short exposures at different LPs for a given mode.

G140L remains at LP3 through Cycle 33 and is planned to migrate entirely to LP11 below LP4 in Cycle 34. The detector area below LP4 has limited cross-dispersion space available for the PSA before reaching the aperture mechanism soft stop, and G140L's relatively narrow profile lets it fit there. The geometric distortion at LP11 is more pronounced than at the central LPs. This is acceptable primarily because G140L is the low-resolution mode and is less sensitive to line spread function distortions than a medium-resolution grating would be, and the improved FUV geometric distortion and walk corrections delivered in 2025 (Indriolo et al. 2025) further reduce this impact. If a medium-resolution grating such as G160M is placed at LP11 in the future (Section \ref{sec:forward}), the more stringent distortion requirements of that mode would be addressed through further refinement of the geometric distortion corrections.

BOA support follows a simpler logic. Because BOA observations are uncommon and bright-target science is generally better served by STIS (Section \ref{subsec:boa}), BOA support under COS2035 is maintained at a small number of LPs rather than commissioned at every new LP. Under current operations, G130M and G160M BOA observations are placed at LP4 and G140L BOA observations are placed at LP3.

\begin{deluxetable}{lllllp{7cm}}
\rotate
 \tabcolsep 4pt
 \tablewidth{0pt}
 \tablecaption{Grating and cenwave assignments under the COS2035 hybrid-LP architecture, including current Cycle 33 assignments and planned LP11 and LP12 commissioning in Cycles 34 and 35. Offsets are measured from LP1 in arcseconds. The Cycle Start column lists the cycle in which the listed assignment took effect or is planned to take effect. \label{tab:assignments}}
 \tabletypesize{\footnotesize}
 \tablehead{
\colhead{LP} & \colhead{Offset} & \colhead{Grating/Cenwave} & \colhead{Wavecal type} & \colhead{Cycle Start} & \colhead{Notes}
 }
 \startdata
 LP3 & $-2.5''$ & G140L, G130M Ly$\alpha$ obs. & TAGFLASH & Cycle 29 & G140L PSA migrates to LP11 (Cycle 34); retains G140L BOA support and G130M Ly$\alpha$ at z=0 \\
 LP4 & $-5.0''$ & G130M, G160M (BOA only) & TAGFLASH & Cycle 25 & BOA support only; no PSA observations \\
 LP5 & $+5.4''$ & G130M/1300,1309,1318,1327 & TAGFLASH & Cycle 29 & Segment A only via COS2025 rules \\
 LP7 & $+8.3''$ & G130M/1055,1096,1222,1291 & SPLIT-wavecal & Cycle 33 & G130M/1291 migrates to LP12 (Cycle 35) \\
 LP10 & $-3.7''$ & G160M & TAGFLASH & Cycle 33 & Hosts both long and short exposures \\
 LP11 & $-6.7''$ & G140L & TAGFLASH & Cycle 34 & Planned; all G140L PSA from LP3 \\
 LP12 & $+1.7''$ & G130M/1291 & TAGFLASH & Cycle 35 & Planned; G130M/1291 from LP7 \\
 \enddata
\end{deluxetable}

\vspace{-0.3cm}
\ssection{LP-infinity and the APT/TRANS rules architecture}\label{sec:framework}

The COS flight software (FSW) maintains tables of operational parameters (aperture block position and focus) with entries for eight lifetime positions. For each cenwave, a lifetime position is defined in the FSW by two values: its aperture block position and its OSM focus setting. For LP1 through LP7, each new lifetime position required loading a fixed set of values into the next available table entry, which in turn required FSW updates, ground system rule updates, and extensive coordination across teams at STScI and Goddard. The commissioning of LP7 and a second lifetime position, planned together for the start of Cycle 33, would have filled the seventh and eighth FSW table entries between them, leaving no room for additional LPs. Two parallel changes were developed concurrently to remove both the FSW table-entry limit and the per-LP cost, the recurring effort required to commission each new position. The LP-infinity framework, described in Section \ref{subsec:lpinfinity}, makes the FSW table entries effectively unlimited from the ground system's perspective and removes the need for FSW updates when adding new LPs. The new architecture in the Astronomer's Proposal Tool (APT) and Transformation (TRANS) rules, described in Section \ref{subsec:apttrans}, replaces cycle-dependent conditional logic with a table-based design that reduces the resources required to add each new LP and minimizes the risk of implementation errors as the rules grow more complex.

\vspace{-0.3cm}
\ssubsection{LP-infinity}\label{subsec:lpinfinity}

LP-infinity is a ground system change rather than a change to the FSW. The eighth entry in the FSW tables, LP8, is permanently reserved as patchable scratch space. The COS and Commanding teams maintain ground system tables of aperture block position and focus values for each commissioned LP. When an exposure is scheduled at any LP beyond LP7, Commanding Instructions are used to set the appropriate parameters for the requested LP in the Science Mission Schedule (SMS). At the start of each exposure, the proper aperture block and focus numbers are loaded into the LP8 locations in the FSW tables. The FSW sees only its eight entries and the values currently loaded; it does not need to know which LP is ``actually'' used. New lifetime positions can therefore be added through ground system table updates alone, with no FSW changes required, as detector real estate and resources permit. LP10 was the first lifetime position commissioned via the LP-infinity patching mechanism, with full LP-infinity support deployed for LP10 calibration starting in late 2025. 

To differentiate the use of FSW entries and patchable tables, the COS team changed the LP naming convention: LP8 is permanently reserved as the patchable scratch entry and is never used as a lifetime position designator. LP9 is skipped, and lifetime positions commissioned after LP7 are numbered LP10 and higher.\footnote{Skipping LP9 ensures that every lifetime position commissioned under the LP-infinity framework carries a double-digit designator (LP10 and higher), visually distinguishing these positions from the original fixed FSW table entries LP1 through LP7.} The LP-infinity framework currently includes placeholder support for LP10 through LP17, meaning that the relevant ground system tables and APT/TRANS structures already accommodate these LP numbers and can be populated with the appropriate values as each LP is commissioned. The framework is straightforwardly extensible beyond LP17 if additional placeholders are required.

\vspace{-0.3cm}
\ssubsection{APT and TRANS rules architecture}\label{subsec:apttrans}

A second piece of the per-LP cost was the complexity of the rules embedded in APT and TRANS that determined which LP applied to a given exposure. APT is the tool used to prepare and submit observing programs, while TRANS, short for Transformation, comprises the rules that translate APT commands into spacecraft commands. Under the architecture in place through Cycle 33, the LP for each exposure was inferred from cascading conditional logic depending on grating, cenwave, segment, FP-POS, cycle, and proposal mode. As the number of supported LPs grew and as different gratings began operating at different LPs simultaneously under the hybrid-LP architecture (Section \ref{sec:hybridlp}), this conditional logic became increasingly difficult to maintain and to test. Starting with Cycle 34, a new table-based architecture replaces this cycle-dependent conditional logic. A single lookup table maps every combination of observation type (science or target acquisition), aperture (PSA or BOA), grating, cenwave, segment, FP-POS, and LP to an allowed-or-disallowed flag, organized along the science/target-acquisition and PSA/BOA axes. APT ingests this table and uses it to validate user selections at the exposure level. TRANS then uses the LP value supplied by APT rather than re-deriving it from internal rules, eliminating the need to update TRANS rules for each new LP. Adding a new LP to this architecture requires only updating the table with the new LP's allowed combinations, a process that takes days rather than weeks.

\begin{figure}[!h]
  \centering
  \includegraphics[width=\textwidth]{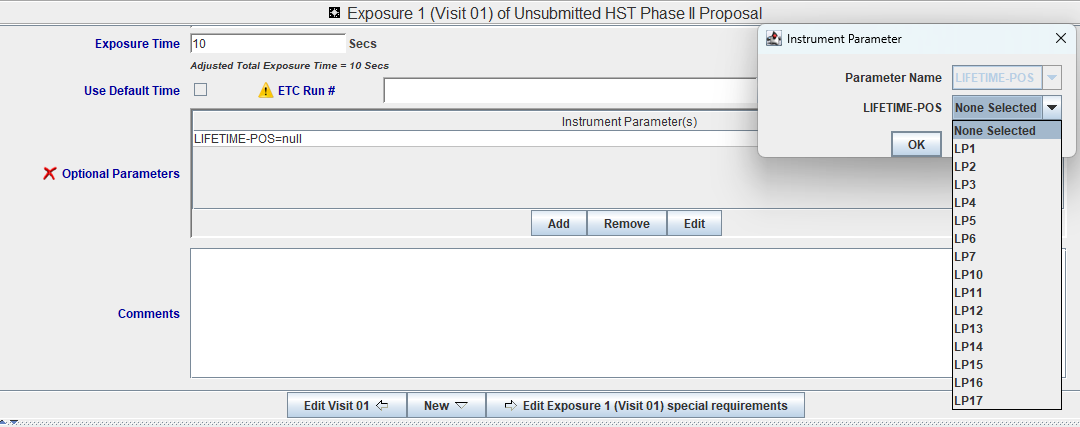}
 \caption{LIFETIME-POS optional parameter selection in APT. Beginning with APT 2026.3 (Phase II for Cycle 34), users specify the LP value at the exposure level for FUV science and target acquisition exposures.}
 \label{fig:apt}
\end{figure}

The user-facing consequence of the new architecture is that LP becomes an explicit optional parameter at the exposure level for FUV science and target acquisition observations, similar in nature to FP-POS or buffer time. Figure \ref{fig:apt} shows the LP selection in APT. To minimize the burden on Phase II preparation and on Contact Scientist review, the rules tables are currently configured so that in Supported Mode each cenwave allows only one LP value. Selecting any other LP returns an error rather than a warning, and the error message lists the LP values valid for that configuration. This ensures that Supported Mode proposals use the appropriate default LP without requiring Contact Scientist intervention. Available (but Unsupported) Mode and Restricted Mode programs may select among additional LPs as the rules tables permit, with the choice subject to the usual Phase I justification requirements. Pre-Cycle-34 programs continue under the existing TRANS logic and do not require reprocessing.

The combination of LP-infinity on the ground system side and the table-based rules architecture in APT and TRANS substantially reduces the per-LP cost of future commissioning. LP-infinity eliminates FSW updates, which had previously been a labor-intensive component of the LP commissioning cycle and required significant work at Goddard. The new APT and TRANS rules architecture eliminates the per-cycle rewrites of conditional logic that had previously consumed substantial APT and TRANS development effort. Future LP commissioning will require coordination across the COS, APT, TRANS, Program Coordinator (PC), and other teams, along with the standard exploratory and enabling work for the LP itself, but at significantly reduced effort and with much shorter implementation timelines.

\vspace{-0.3cm}
\ssection{Revised gain-sag flagging method}\label{sec:gainsag}

The COS FUV detector is subject to gain sag, the gradual decline in microchannel plate response as a function of accumulated charge at each pixel location. As the modal gain at a pixel falls, photon events can no longer be reliably centroided and the apparent sensitivity at that pixel drops. The CalCOS pipeline tracks gain-sagged pixels through the GSAGTAB reference file, which lists pixels to be excluded from spectral extraction. CalCOS originally used a BOXCAR extraction algorithm in which flagging any pixel within the extraction region rejected the entire column. The two-zone extraction method (Proffitt 2015) reduced this aggressiveness by separating the inner and outer cross-dispersion contours of the spectral profile, so that GSAGTAB flags in outer-contour pixels no longer trigger column rejection. Before Cycle 33, a pixel was added to the GSAGTAB whenever its modal gain fell to 3, corresponding to approximately 5\% count loss in that pixel. Even under two-zone extraction, this per-pixel threshold remained conservative: flagging a single inner-contour pixel removed the entire column from the extraction, even though that pixel was typically one of many contributing to the column and its absence had a negligible effect on the integrated flux.

\begin{figure}[!h]
  \centering
\includegraphics[width=0.93\textwidth]{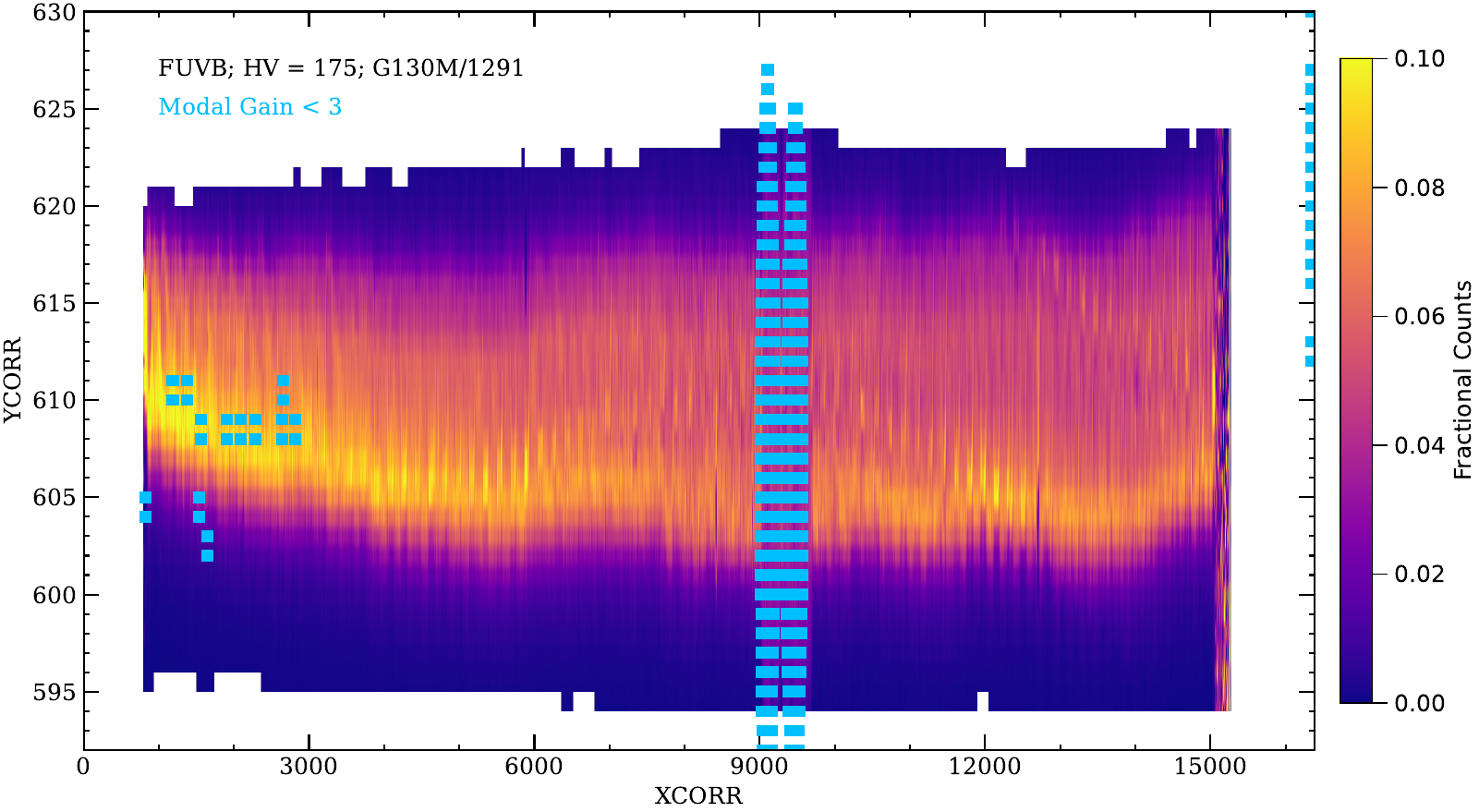}\\[2mm]
  \includegraphics[width=0.93\textwidth]{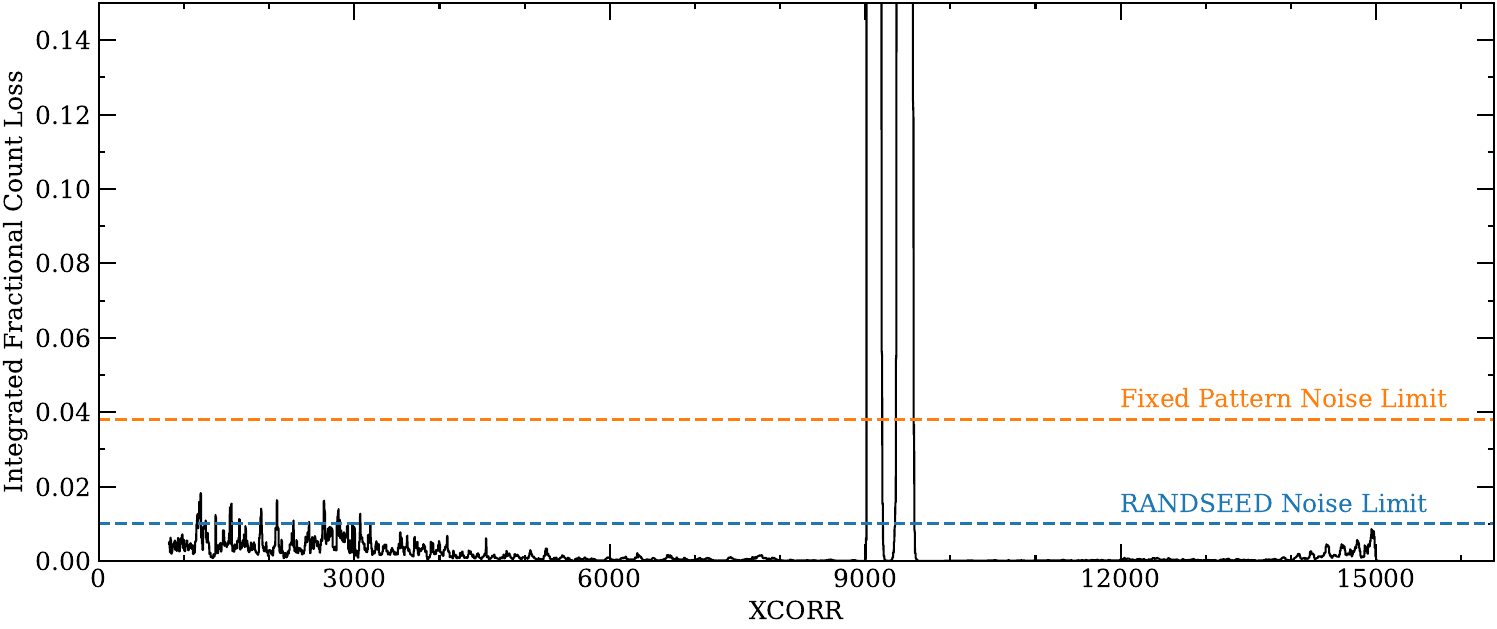}
 \caption{Top: fractional count map for G130M/1291 on the FUVB segment, showing the contribution of each pixel to the integrated flux along its column, with pixels reaching modal gain 3 highlighted. These highlighted pixels include both the Ly$\alpha$ airglow holes and the continuum-driven gain sag on the short-wavelength side of the segment. Bottom: integrated count loss along each XCORR pixel within the extraction region. The large count-loss spikes correspond to the Ly$\alpha$ airglow columns, where the integrated loss along the full column is high; these columns remain flagged under the existing rules. The continuum-driven modal-gain-3 pixels relevant to the revised criterion produce integrated count losses well below the maximum achievable S/N for G130M/1291 set by fixed-pattern noise, indicating that the previous per-pixel flagging method was overly conservative for continuum-dominated gain sag. The revised flagging method flags only pixels whose contribution to integrated column count loss exceeds the maximum-S/N threshold, so the airglow columns remain flagged while the continuum sag below threshold does not.}
 \label{fig:gainsag}
\end{figure}

The COS2025 rules (Oliveira et al. 2018) concentrated airglow exposure onto a small set of detector locations intentionally allowed to sag, with the surrounding regions protected from airglow damage. In the post-COS2025 era, continuum exposure rather than airglow is therefore the dominant driver of new gain sag, and continuum sag affects spectral extraction differently from airglow sag. Airglow lines fill the aperture and depress the modal gain of many pixels along the column, so when a column reaches modal gain 3 from airglow the integrated count loss along the column is substantial and full-column rejection is appropriate. Continuum exposure depresses only those pixels that fall on the spectral profile, so only a subset of the pixels along a given column reach modal gain 3. For G130M/1291 at LP5 in 2025, a handful of pixels on the short-wavelength side of FUVB reached modal gain 3 within the inner zone of the two-zone extraction. These pixels triggered full-column rejection under the per-pixel threshold, even though the integrated count loss along each affected column was below the RANDSEED noise level of approximately 1\% and well below the maximum achievable S/N for G130M/1291 set by fixed-pattern noise, approximately 3.8\% (Figure \ref{fig:gainsag}).

The revised flagging method introduced in 2025 takes the integrated column behavior into account (Payne \& Dixon 2026, Section 5.12.2). A pixel with modal gain below 3 is added to the GSAGTAB only when its contribution to the integrated count loss along its column would exceed the maximum achievable S/N for the mode in use, with the maximum-S/N threshold drawn from the 50th-percentile values for each grating and FP-POS combination tabulated in Roman-Duval et al. (2023). That contribution is weighted by the cross-dispersion profile of the mode, which is derived from the 2D projection of high S/N continuum source spectra. Airglow-dominated gain sag continues to be flagged as before. Continuum-dominated gain sag below the S/N threshold is no longer flagged, and testing on representative archival data confirmed that the resulting X1DSUM spectra are nearly indistinguishable from those that would have been produced without any gain sag at the affected pixels. The change has been deployed in the CalCOS pipeline through updated GSAGTAB reference files.

The operational benefits are twofold. First, the revised method extends the viable lifetime of every active LP, since pixels that would previously have triggered column rejection no longer do so when they contribute negligibly to the total spectrum. Second, it makes LP12 viable: the detector region between LP1 and LP2 (Figure \ref{fig:lpmap}) has accumulated gain sag from earlier LP1 and LP2 operations, and under the previous flagging method the residual sag would have been flagged aggressively in any G130M/1291 spectrum extracted there. Under the revised method, most of the residual sag in that region contributes a count loss below the S/N threshold for G130M/1291 and does not affect the extracted spectrum, allowing G130M/1291 to operate at LP12 with adequate spectral quality (Section \ref{sec:forward} describes the LP12 commissioning). The same fixed-pattern-noise S/N threshold from Roman-Duval et al. (2023) also informs the new maximum-S/N usage policy described in Section \ref{subsec:maxsn}.

\vspace{-0.3cm}
\ssection{The COS2035 strategy}\label{sec:strategy}

The COS2025 strategy (Oliveira et al. 2018) was introduced in 2017 to prolong the FUV detector lifetime through 2025 by isolating airglow exposure to a small set of detector locations and concentrating gain sag there rather than spreading it across the detector. The COS2025 rules implement this strategy at the cenwave level. G130M cenwaves at or above 1300\,\AA{} (1300, 1309, 1318, and 1327) may not be used with both segments on; only Segment A operation is permitted under the standard rules. To limit the number of gain-sag holes from Ly$\alpha$ airglow on FUVB, G130M/1291 with both segments on is restricted to FP-POS=3 and 4; FP-POS=1 and 2 are available only with Segment A. The blue G130M cenwaves (1055 and 1096) and G130M/1222 carry no segment or FP-POS restrictions; G160M and G140L are unrestricted,  though G140L cenwaves 800 and 1105 use only Segment A by default. To enable continued observations of Ly$\alpha$ at zero redshift, G130M/1309, 1318, and 1327 may be used at LP3 with both segments on. These rules remain in force under COS2035 and continue to govern how the FUV detector accumulates counts.

The COS2035 strategy extends COS2025 in two respects. First, the four technical breakthroughs described in Sections \ref{sec:splitwave} through \ref{sec:gainsag} (SPLIT-wavecals and the new detector real estate, the hybrid-LP architecture, the LP-infinity ground system framework, and the revised gain-sag flagging method) together open access to detector area and operational flexibility that COS2025 alone could not provide. Second, two new usage policies, introduced in Cycles 32 and 33 and described in Sections \ref{subsec:maxsn} and \ref{subsec:twopercent}, cap per-target S/N and per-program lifetime usage in response to higher-than-historical detector usage in recent cycles. Together, the COS2025 rules, the four technical breakthroughs, and the two new usage policies constitute the COS2035 strategy.

\vspace{-0.3cm}
\ssubsection{Maximum achievable signal-to-noise per target}\label{subsec:maxsn}

Each combination of grating and number of FP-POS observations on the COS FUV detector has a maximum achievable S/N ratio set by fixed-pattern noise. Above this S/N value, longer exposures no longer add usable signal but continue to deplete the detector lifetime at the LP in question. To preserve detector lifetime in the face of higher-than-historical usage, COS observations should not exceed the maximum achievable S/N for the grating and FP-POS combination in use. The 50th-percentile maximum-S/N values for each combination are tabulated in Table 1 of Roman-Duval et al. 2023 (COS Instrument Science Report 2023-11). Exceptions are permitted for specific science cases as justified in Phase I. This policy was introduced in Cycle 32 and is documented in the \href{https://www.stsci.edu/contents/news/cos-stans/july-2024-stan.html\#BrightObjects}{July 2024 STScI STAN}.

\vspace{-0.3cm}
\ssubsection{Two-percent per-program lifetime usage cap at any single LP}\label{subsec:twopercent}

A pixel on the COS FUV detector can accumulate approximately 27,000 counts before reduced gain at that location results in significant flux loss. Two percent of this lifetime corresponds to 540 counts in the brightest pixel. Under the COS2035 policy introduced in Cycle 33, no single observing program may deposit more than 540 counts in any pixel at any single LP,\footnote{Counts from geocoronal Lyman-$\alpha$ airglow are exempt from this limit.} integrated over all targets and exposures in the program. Programs whose anticipated usage approaches this cap, within a factor of two (i.e., above 1\% of the pixel lifetime, or 270 counts), must verify their compliance and justify their lifetime usage in the Phase I proposal. This policy is documented in the \href{https://www.stsci.edu/contents/news/cos-stans/march-2025-stan.html\#PolicyChangesMaxLifetime}{March 2025 STScI STAN}.

Table \ref{tab:lifetime} provides a quick-check estimate of the maximum number of targets a program can observe before reaching the 2\% cap, as a function of grating, cenwave, segment, target S/N, and LP. The assignments shown reflect the Cycle 34 and 35 defaults, with G140L at LP11 and G130M/1291 at LP12; under the Cycle 33 assignments these modes were at LP3 and LP7 respectively. The $\lambda_{\rm SNR}$ column lists the wavelength at which the target S/N is evaluated. The blue modes G130M/1055 and 1096 on Segment B support relatively few targets because reaching the target S/N at the low-throughput blue end requires long exposures, and those exposures drive the brightest pixel toward the per-pixel cap on Segment A, where the source is typically brighter and the throughput higher. One mitigation is to turn off the higher sensitivity Segment A for part or all of the observations. Programs whose anticipated lifetime usage approaches the 2\% cap, within a factor of two (above 1\% of the pixel lifetime), should perform a detailed ETC-based calculation rather than relying on the table alone. The detailed calculation requires running the ETC for each target to determine its brightest-pixel count rate, then multiplying by the appropriate exposure time and summing across visits to obtain the total counts deposited at the LP. For continuum-dominated observations, the brightest pixel sees the same continuum regardless of FP-POS, so the relevant exposure time is the total per grating per visit. For emission-line-dominated observations, the bright emission lines fall on different pixel locations at different FP-POS settings, so the relevant exposure time is the total per FP-POS per visit. Programs with multiple visits at the same LP sum the counts deposited in each visit to obtain the total lifetime impact at that LP. If the result exceeds 540 counts, programs must adjust by selecting a less sensitive mode, turning off a segment, reducing exposure time, or moving observations to STIS.

Three worked examples illustrate the procedure. (1) A bright A3V star observed with G160M for 11,590\,s deposits approximately 1,500 counts in the brightest pixel, exceeds the 540-count cap by a factor of approximately 2.78, and requires an alternative such as STIS/E140M. (2) Epsilon Eridani observed with G130M/1222 for 12,000\,s using four FP-POS deposits approximately 213 counts per FP-POS in the brightest pixel, well below the 540-count cap and below the 270-count justification trigger. (3) A program intending to observe 40 white dwarfs with G130M/1291 at S/N=27 exceeds the Table \ref{tab:lifetime} capacity of 22 targets at that combination and requires either a less-impactful grating (e.g., G130M/1222) or moving some observations to STIS. Detailed worked examples with ETC calculation references are provided in the \href{https://www.stsci.edu/contents/news/cos-stans/march-2025-stan.html\#PolicyChangesMaxLifetime}{March 2025 STScI STAN}.

\begin{deluxetable}{lcccccccc}
 \tabcolsep 4pt
 \tablewidth{0pt}
 \tablecaption{Estimated number of targets before a program will hit 2\% lifetime usage as a function of mode, target S/N, and segment. Users are encouraged to do more careful calculations if they have several targets or high S/N requirements. Values for G130M/1291 at LP12 and G140L at LP11 are preliminary pending ETC support at those positions.\label{tab:lifetime}}
 \tabletypesize{\footnotesize}
 \tablehead{
\colhead{Grating} & \colhead{CENWAVE} & \colhead{LP\#} & \colhead{Moderate} & \colhead{Maximum} & \colhead{Segment} & \colhead{$\lambda_{\rm SNR}$} & \colhead{\# Targets} & \colhead{\# Targets} \\
\colhead{} & \colhead{} & \colhead{} & \colhead{SNR} & \colhead{SNR} & \colhead{} & \colhead{} & \colhead{mod. SNR} & \colhead{high SNR}
 }
 \startdata
 G130M & 1055 & 7 & 10 & --& B& 1030 & 6& --  \\
 G130M & 1055 & 7 & 10 & 30& A& 1100 & 60  & 8\\
 G130M & 1096 & 7 & 10 & --& B& 1060 & 6& --  \\
 G130M & 1096 & 7 & 10 & 30& A& 1120 & 280 & 30  \\
 G130M & 1222 & 7 & 10 & 27& B/A & 1230 & 520 & 80  \\
 G130M & 1291 & 12 & 10 & 27& B/A & 1310 & 200 & 22  \\
 G130M & 1327 & 5 & 10 & 27& A& 1340 & 230 & 30  \\
 G160M & 1533 & 10 & 10 & 35  & B/A & 1540 & 110 & 8\\
 G160M & 1623 & 10 & 10 & 35  & B/A & 1635 & 60  & 4\\
 G140L & 800  & 11 & 10 & 36& A& 1300 & 190 & 14  \\
 G140L & 1280 & 11 & 10 & 36& A& 1300 & 200 & 14  \\
 \enddata
\end{deluxetable}
\vspace{-0.3cm}
\ssection{Looking Forward: LP11, LP12, and beyond}\label{sec:forward}

\begin{figure}[!h]
  \centering
  \includegraphics[width=\textwidth]{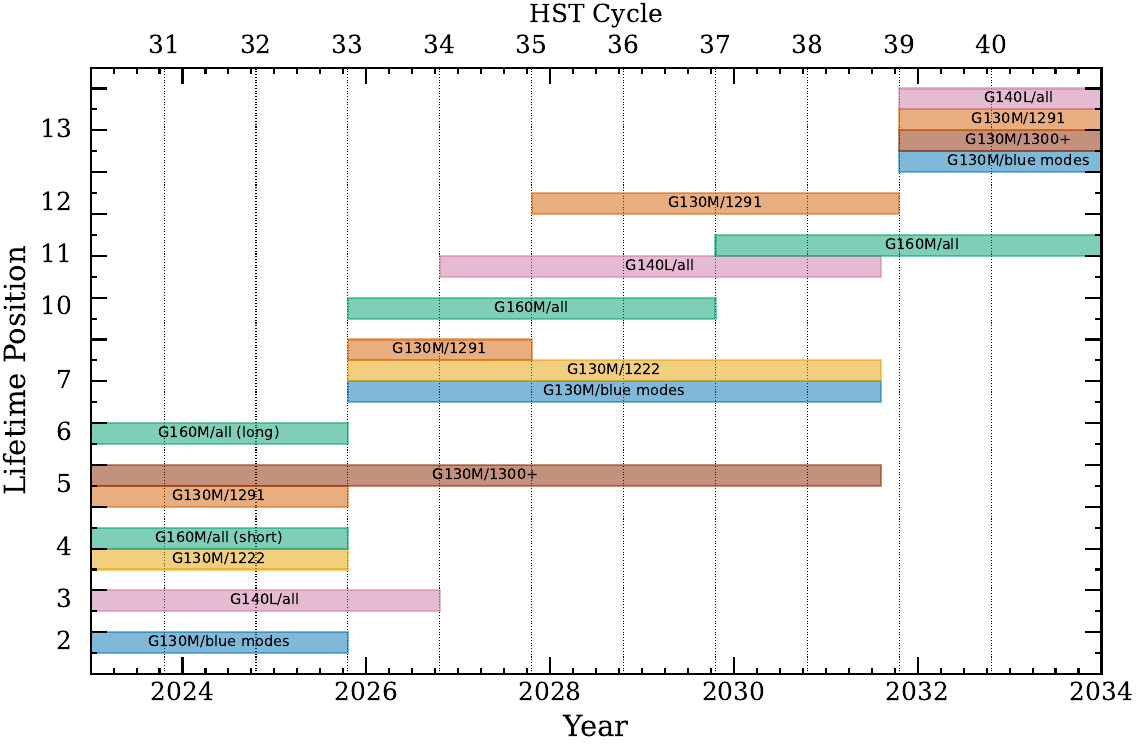}
 \caption{Projected assignment of FUV gratings and cenwaves to lifetime positions through the early 2030s. Each colored bar shows the operational period of a given grating/cenwave at the indicated lifetime position. LP11 and LP12 are in commissioning for Cycles 34 and 35 respectively. Placements beyond LP12 (Cycle 36 and later), including the G160M migration and any subsequent lifetime positions, are notional: they are based on current detector usage models, represent one of several candidate scenarios, and are expected to evolve. Timelines for LP12 and above are also subject to funds and resource availability. Additional lifetime positions and strategies are available to extend COS operations into the 2040s, well beyond the timeframe shown here.}
 \label{fig:timeline}
\end{figure}

The COS2035 strategy is structured to extend FUV operations through the 2030s and beyond. Two new lifetime positions are in active commissioning (LP11 below LP4 and LP12 between LP1 and LP2) with additional detector area above LP7 reserved for further LPs as future needs arise. Figure \ref{fig:timeline} shows the projected assignment of each grating and cenwave to its LP through the early 2030s.

LP11 is being commissioned for science use beginning with Cycle 34 in November 2026. LP11 will host all G140L cenwaves, migrated entirely from LP3, with placement below LP4 in the detector area where geometric distortion is more pronounced but acceptable for the low-resolution G140L mode (Section \ref{sec:hybridlp}). LP11 operates under TAGFLASH and incurs no SPLIT-wavecal overhead, since it will sit below the $+5.5''$ light leak.

LP12 is being commissioned for science use beginning with Cycle 35. LP12 will host G130M/1291, migrated entirely from LP7, with placement between LP1 and LP2 to provide high spectral resolution and TAGFLASH operation for the most heavily used FUV mode (Section \ref{sec:hybridlp}). LP12 is viable because the revised gain-sag flagging method (Section \ref{sec:gainsag}) tolerates the residual gain sag accumulated from earlier LP1 and LP2 operations: that sag is concentrated in the outer wings of the cross-dispersion profile and does not significantly affect the flux accuracy of G130M/1291 spectra extracted there, with the integrated count loss below the maximum S/N achievable for the mode set by fixed-pattern noise.

The G160M modes currently at LP10 will need to migrate next after LP12, probably before any of the G130M modes, and two placement options are under consideration to avoid an increase in SPLIT-wavecal overhead. The first places G160M at LP11 alongside G140L, accepting the lower resolution and more pronounced geometric distortion below LP4 in exchange for keeping G160M below the $+5.5''$ light leak; the narrow cross-dispersion profiles of the two modes separate well at LP11, and the high-voltage settings can be lower for G160M since it would sit farther from LP4 than G140L. This option is reflected in the LP timeline of Figure \ref{fig:timeline}, with the G160M migration to LP11 projected for the early 2030s. The second option places G160M between LP1 and LP3, where the narrower LP1 to LP3 separation and airglow on both sides of the profile make placement more demanding, but the narrow cross-dispersion footprint of G160M may allow it to fit with the new gain-sag flagging methodology (Section \ref{sec:gainsag}) from COS2035. This second option would provide higher spectral resolution and reduced geometric distortion so would be preferred if it fits. Selecting between these options will require careful gain modeling and the chosen placement will be communicated in the future. 

Detector area above LP7 remains available for additional lifetime positions beyond LP12 (these would require SPLIT-wavecals, since they sit above the light leak), which is why lower real estate below the light leak is used first when possible. Area at the bottom of the detector remains available for a BOA-supporting lifetime position should one be needed. The modes currently hosted at LP7 and LP10 will need to migrate as detector conditions at those LPs evolve over the early 2030s; depending on detector conditions and usage patterns at that time, they could be consolidated at a single new lifetime position above LP7 or split across multiple new LPs. Beyond 2035, other operational approaches are available to ensure that high-sensitivity medium-resolution FUV spectroscopy remains possible through the next decade.

\vspace{-0.3cm}
\ssection{Summary and conclusions}\label{sec:summary}

The COS2035 strategy extends FUV operations through the 2030s and beyond by combining the COS2025 rules (Oliveira et al. 2018), which remain in force, with four technical breakthroughs and two new usage policies. The four technical breakthroughs are SPLIT-wavecals and the new detector real estate above the $+5.5''$ light leak (Section \ref{sec:splitwave}), the hybrid-LP architecture in which different gratings and cenwaves operate at different LPs simultaneously (Section \ref{sec:hybridlp}), the LP-infinity ground system framework along with its supporting APT and TRANS rules architecture (Section \ref{sec:framework}), and the revised gain-sag flagging method that evaluates integrated column count loss against the maximum achievable S/N per mode (Section \ref{sec:gainsag}). The two new usage policies, layered on top of COS2025, cap per-target S/N at the maximum achievable value set by fixed-pattern noise (Section \ref{subsec:maxsn}) and limit any single program to no more than 2\% of the lifetime at any single LP (Section \ref{subsec:twopercent}). Together, these elements constitute the COS2035 strategy.

With the LP7 and LP10 commissioning completed at the start of Cycle 33 in November 2025, the COS team has implemented eight lifetime positions over the nearly 17-year operational history of the instrument. LP11 and LP12 are in active commissioning for Cycles 34 and 35 respectively, and detector area above LP7 remains available for additional LPs as future needs arise. Given the excellent health of both HST and the COS instrument, the COS2035 strategy is intended to extend the scientific productivity of Hubble's unique UV capabilities as far into the future as possible. In doing so, it helps bridge the gap to the Habitable Worlds Observatory (HWO) and enables concurrent UV spectroscopic observations with JWST and Roman over the intervening years.

For current operational details, including the most up-to-date LP assignments, observing modes available, and usage policy specifics, COS users should consult the HST Instrument Handbook for COS, the Call for Proposals for the relevant cycle, and the COS Space Telescope Analysis Newsletters at \url{https://www.stsci.edu/contents/news/cos-stans/}. New LP assignments and operational guidance will continue to be communicated through these channels as detector conditions evolve.

\vspace{-0.3cm}
\ssectionstar{Acknowledgements}
\vspace{-0.3cm}
{\it We dedicate this ISR to the memory of William (Will) J. Fischer, whose contributions helped extend the life of COS. He is deeply missed.}\\

\noindent We thank the COS team at STScI and the COS Flight Operations Team at NASA Goddard Space Flight Center for the work that made the COS2035 strategy possible. We also thank the APT, TRANS, Commanding, SOC SE, PDB, SCIOPSDB, and Program Coordinator teams for their contributions to the rules architecture and operational support. 

\vspace{-0.3cm}
\ssectionstar{Change History for COS ISR 2026-04}\label{sec:History}
\vspace{-0.3cm}
Version 1: \ddmonthyyyy\today\ - Original Document

\vspace{-0.3cm}
\ssectionstar{References}\label{sec:References}
\vspace{-0.3cm}

\noindent
Dos Santos, L. A., et al.\ 2025, COS Instrument Science Report 2025-13
\\
Fox, A. J., et al.\ 2020, COS Instrument Science Report 2020-05
\\
Indriolo, N., et al.\ 2025, COS Instrument Science Report 2025-07
\\
James, B. L., et al.\ 2023a, COS Instrument Science Report 2023-10
\\
James, B. L., et al.\ 2023b, COS Instrument Science Report 2023-15
\\
Morse, J. 1999, COS Aperture Plate Baseline Design, COS-11-0009
\\
Oliveira, C., et al.\ 2013, COS Instrument Science Report 2013-02
\\
Oliveira, C., et al.\ 2018, COS Instrument Science Report 2018-16
\\
Osten, R., et al.\ 2013, COS Instrument Science Report 2013-16
\\
Payne, A., \& Dixon, W., eds.\ 2026, COS Instrument Handbook, Version 18.0
\\
Plesha, R., et al.\ 2018, COS Instrument Science Report 2018-23
\\
Proffitt, C.\ 2015, COS Instrument Science Report 2015-03
\\
Roman-Duval, J., et al.\ 2016, COS Instrument Science Report 2016-01
\\
Roman-Duval, J., et al.\ 2018, COS Instrument Science Report 2018-14
\\
Roman-Duval, J., et al.\ 2023, COS Instrument Science Report 2023-11
\\
Rowlands, K., et al.\ 2024, COS Instrument Science Report 2024-08
\\
Sahnow, D., et al.\ 2021, Proc. SPIE, 11821, 118211H
\\
STScI 2024, COS Space Telescope Analysis Newsletter,
\href{https://www.stsci.edu/contents/news/cos-stans/july-2024-stan.html\#BrightObjects}{July 2024}
\\
STScI 2025, COS Space Telescope Analysis Newsletter,
\href{https://www.stsci.edu/contents/news/cos-stans/march-2025-stan.html\#PolicyChangesMaxLifetime}{March 2025}

\end{document}